\documentclass[eqsecnum,aps,showpacs]{revtex4}

\def\beq{\begin{equation}}
\def\eeq{\end{equation}}

\newcommand{\gsim}{\raisebox{-0.8ex}{\mbox{$\stackrel{\textstyle >}{\sim}$}}}

\begin{document}

\title{Optimum Placement of Post-1PN GW Chirp Templates\\
Made Simple  at any Match Level\\
via Tanaka-Tagoshi Coordinates}
\author{R.P. Croce}
\affiliation{Wavesgroup, D.I.$^{3}$E., University of Salerno, Italy}
\vspace*{-1cm}
\author{Th. Demma, V. Pierro and I.M. Pinto} 
\affiliation{Wavesgroup, University of Sannio at Benevento, Italy}
\date{\today}


\begin{abstract}
A simple recipe is given for constructing a maximally sparse
regular lattice of spin-free post-1PN gravitational wave chirp templates
under a given minimal-match constraint,
using Tanaka-Tagoshi coordinates. Cardinal interpolation
among the resulting maximally-sparse templates  is discussed.
\end{abstract}


\maketitle

\section{Introduction}

~~Gravitational wave  {\em chirps}
emitted by compact binary stars in their adiabatic inspiral phase 
will be primary targets for the early operation of
broadband  laser-interferometric detectors like
TAMA300, GEO600, LIGO and VIRGO  \cite{sources}.
These signals have been thoroughly studied and accurately modeled \cite{signals}.

The maximum-likelihood strategy \cite{Helstrom}
for detecting signals of known shape
(except for a set of unknown parameters)
in additive stationary colored gaussian noise
consists in correlating the data
with a set ${\cal L} = \{ g_i \}$
of possible expected waveforms ({\em templates}),
and using the largest correlator as a detection  statistic \cite{Helstrom}. 
The mentioned correlators are the (noise-weighted)
scalar products:
\beq
\langle
a,g
\rangle=
2\left[
\int_{f_i}^{f_s}
a(f)g^*(f)
\frac{df}{\Pi(f)}
+c.c.
\right],
\label{eq:correlator}
\eeq
where $(f_i,f_s)$ is the antenna spectral window,
$a(f)\!=\!h(f)\!+\!n(f)$ are the noise-corrupted (spectral)
data, $h(f)$ is a (possibly null) signal, 
$n(f)$ is a realization of the antenna noise 
with (one-sided) power spectral density $\Pi(f)$,
$g(f)$ is a template, 
and $c.c.$ denotes complex-conjugation.

The issue of optimum  placement of the templates 
in the waveform parameter-space is more or less obviously 
a crucial  and still open one,
and has been addressed by several Authors \cite{Sat_Dhu}-\cite{Owe_Sat}.

At fixed false-alarm probability, the detection probability
is an increasing function of the {\em overlap} 
\beq
{\cal O}(h,g)=
\frac{\langle h,g \rangle}
{||h||\cdot||g||} \leq 1
\label{eq:overlap}
\eeq
between the signal $h$ and the template $g$,
where the norm $||u||=\langle u,u \rangle^{1/2}$.
The template set ${\cal L}$ should be designed in such a way
that, for any admissible signal $h$,
the overlap never drops below  a prescribed value $\Gamma$,
which can be immediately related 
to the fraction $(1-\Gamma^3)$ of potentially observable sources
which might be missed out \cite{Apo}. 

The overlap can be readily maximized w.r.t. the (unknown but irrelevant)
initial-phase and coalescency-time differences 
between the signal and the template \cite{partialmax}. 
The resulting partially-maximized overlap 
is called the {\em match}, is denoted by $M(h,g)$,  and is a function 
of the source and template {\em intrinsic} parameters only
(e.g., at 2.5PN order, the binary companion-masses, spin-spin and spin-orbit parameters).
The minimum-overlap condition rephrases obviously 
into the following minimal-match prescription:
\beq
\forall h \in {\cal S},~~
\exists g \in {\cal L}~:~
M(h,g) \geq \Gamma,
\label{eq:min_mat}
\eeq
where ${\cal S}$ is the set of admissible signals.


Discussion will be  henceforth restricted to the $2D$
parameter-space of spin-free chirp-waveforms,
in view of the  widespread present consensus
that these should be adequate 
for the early operation of interferometric detectors \cite{Owe_Sat}. 

In the following we shall denote as $G$ 
the  (2D, spin-free) parameter-space point 
corresponding to waveform $g$.
We shall also denote as $\gamma_{\Gamma}(G)$ 
the match contour-line, whose points
represent the set of waveforms $h$ such that  $M(h,g)=\Gamma$,  
and as $S_{\Gamma}(G)$ the 2D-region bounded by $\gamma_{\Gamma}(G)$, 
representing the set of waveforms $h$ such that $M(h,g) \geq \Gamma$.

The template set $\Lambda=\{G_i\}$ should fulfill 
the following basic requirements.
First of all,  {\bf (a)}  the minimal-match condition (\ref{eq:min_mat}) 
should be met, which re-phrases into
\beq
\bigcup_{i} S_{\Gamma}(G_i) \supseteq \Sigma,
\label{eq:min_mat0}
\eeq
where $\Sigma$ is the image of ${\cal S}$ i.e., the 
subset of the waveform parameter-space 
corresponding to admissible sources.
Pictorially, (\ref{eq:min_mat0}) means that the patches
$S_{\Gamma}(G_i)$  should cover the whole set $\Sigma$  without leaving holes.
At the same time, {\bf (b)} the template set $\Lambda$  should be chosen 
{\em as much sparse as possible}, so as to minimize the overall 
number of templates, and hence the detection threshold-level, 
for any prescribed false-alarm probability \cite{Helstrom}.

It might be further desirable {\bf (c)}  to deal 
with a set $\Lambda$  forming  a  regular
(or piecewise-regular) grid, for computational ease.

A further template-placement requirement  is also worth mentioning.
When covering $\Sigma$ with  (regularly spaced or patch-wise regularly spaced)
templates, a certain amount of template spill-over 
beyond $\partial\Sigma$ is unavoidable. 
The post-1PN spin-free  waveform  parameter-space subset $\Sigma$
corresponding to admissible sources, 
for which  $m_{min}\leq m_1\leq m_2 \leq m_{max}$,
is a three-vertex curved-side $2D$ domain.
Spill-over across the $m_1\!=\!m_2$ boundary line of  $\Sigma$  
(equal-mass boundary-line)  is  a true {\em penalty}, and thus {\bf (d)}
should be kept to a minimum.
This is not the case  for the other two boundary lines of $\Sigma$, 
where points beyond $\partial\Sigma$
could well correspond to {\em possible} (though unlikely) sources.

At {\em large} values of the minimal match, 
e.g. typically $M(h,g) \geq  \Gamma~\gsim~0.97$,
as prescribed for a single-step search strategy \cite{Owe_Sat},
the match is very well approximated by a {\em quadratic} function 
\cite{Owen} of the distance 
between  $H$ and $G$ in parameter-space,
and the (approximate) match contour-lines $\gamma_{\Gamma}(G)$  are ellipses.
However when using  post-1PN order waveforms 
(which will be required in a one-step search 
to achieve the prescribed {\em large} minimal-match levels \cite{Apo}),
as $G$ moves throughout  $\Sigma$,
the above ellipses  rotate and stretch 
as an effect of the {\em intrinsic} nonzero curvature of $\Sigma$.

An  effective procedure for constructing a $2D$ spin-free
post-1PN template set appropriate to this case has been formulated in \cite{Owe_Sat}, 
as a generalization of  the 1PN strategy introduced in  \cite{Owen}. 
The simplest  rectangular-cell template-lattice is chosen,
and the span of template $G_i$ 
(i.e., the set of points representing waveforms
for which $M(h,g_i) > M(h,g_j), \forall j \neq i$)
is accordingly taken as the largest rectangle inscribed 
in the minimal-match contour-line $\gamma_{\Gamma}(G_i)$. 
In order to minimize spill-over across the equal-mass 
boundary-line of  $\Sigma$, while optimizing template
placement for equal-mass sources 
(which are credited to be most abundant \cite{Taylor}),
the algorithm is started by placing a sufficient number of templates
along the equal-mass boundary line. 
The nearest-neighboring templates  are  placed in such a way
that the boundaries of their rectangular spans  {\em touch}
without intersecting.
As a result $\Sigma$ is covered with a set of patches,
each of which is a stack of equal-width rectangles, whose
height changes according to the shear/stretch of the 
elliptical $\gamma_{\Gamma}$ contour-lines.
The current release of GRASP \cite{GRASP}
implements a similar procedure,  
resulting into a relatively straightforward template placement algorithm.

As already noted in \cite{Owe_Sat} the above algorithm is
{\em not} optimal, i.e., it  does not necessarily  satisfy requirement {\bf (b)},
in view of its {\em a-priori} restriction to rectangular cells \cite{falsealarm}.

The template placement issue becomes even more complicated
at low minimal-match levels (e.g., typically  $\Gamma \gsim 0.7$) 
as prescribed in the first step(s) of  hierarchical search strategies.
In multi-step hierarchical searches 
a {\em single} template-lattice $\Lambda$ 
designed to achieve a {\em large} $\Gamma$ throughout  $\Sigma$ is constructed.
In the first step only  the correlators corresponding to
a  suitably {\em decimated}  subset  of  $\Lambda$  are computed,
covering  $\Sigma$  at a  {\em lower}  minimal-match level.
In the second step, only  a limited number of correlators,
corresponding to a few non-decimated subsets  of $\Lambda$
need to be actually computed, covering
suitable neighbourhood(s)  of the candidate signals 
discovered in the first step.
Multi-step hierarchical search strategies
should allow a sizeable reduction in the required
computing power \cite{MoDhu}-\cite{alternative}.
Mohanty \cite{Mohanty} formulated a simple strategy 
for evaluating the lattice decimation-steps 
(taken as constant  throughout extended patches of the lattice)
for a post-1PN template lattice.
Also in this case, however, requirement {\bf (b)} will
be most certainly violated. 
By itself,  the template placement issue at low minimal-match 
levels is further complicated by the fact that the
match contour-lines $\gamma_{\Gamma}(G)$ besides
varying with $G$ are {\em no longer} elliptical.
For this reason, the procedure expounded in \cite{Mohanty}
can only be validated by extensive Monte-Carlo trials.

In this rapid communication we present a possible way to get  rid of all the above 
difficulties and limitations, by capitalizing on the remarkable properties
of the post-1PN waveform parametrization
introduced by  Tanaka and Tagoshi \cite{TaTa},
and further discussed in \cite{TaTa1}.
We accordingly  introduce a simple and systematic procedure
to set up an optimum template-lattice \cite{lattice}, 
which closely  fulfills {\em all}  conditions {\bf a)}, {\bf b)} and
{\bf c)} above,  at  {\em any} prescribed minimal-match level \cite{spill_over}.

The paper is organized as follows.
In Sect. 2  the Tanaka-Tagoshi parametrization  and its relevant
properties, including the main features of the match contour-lines,
are briefly introduced.  
In Sect. 3  our optimum template placement procedure is exposed in detail.
In Sect. 4  the gain achievable by cardinal interpolation 
among these optimally-placed templates is discussed.
Conclusions follow under Section 5.

\section{Tanaka-Tagoshi Parametrization of Post-1PN Waveforms}

~The Tanaka-Tagoshi transformation \cite{TaTa2}
is a linear one-to-one mapping 
$F \in \Sigma \leftrightarrow \tilde{F} \in {\cal T}$
between the (2D, spin-free, post-1PN)
admissible waveform parameter-space $\Sigma$
and a {\em globally flat}  manifold ${\cal T}$ 
such that:  i) the ratio
\beq
\epsilon = 
\left|
\frac{M(h,g)-M(\tilde{h},\tilde{g})}{M(h,g)}
\right|
\label{eq:match_error} 
\eeq
is {\em negligibly small} (typically of the order of $10^{-3}$ \cite{TaTa1})
and, ii)  $M(\tilde{h},\tilde{g})$  depends {\em only}  on the
{\em distance} between $\tilde{H}$ and $\tilde{G}$ \cite{TaTa}.
The Tanaka-Tagoshi construction is 
effective for all first generation interferometers \cite{TaTa1}. 

Clearly, property  i) above allows to solve the optimum template placement
problem in the {\em globally flat}  manifold ${\cal T}$, by
finding  a  set  $\tilde{\Lambda}=\{\tilde{G}_i\} \subset {\cal T}$  such that:
\beq
\bigcup_{i} S_{\Gamma}(\tilde{G}_i) \supseteq {\cal T}.
\label{eq:min_mat1}
\eeq
On the other hand, property  ii)  implies that  the sets $S_{\Gamma}(\tilde{G}_i)$
are {\em identical}  up to trivial translations.
This means that $\tilde{\Lambda} \subset {\cal T}$
is a {\em globally uniform} template lattice,
featuring a {\em regular grid} of nodes, 
at {\em any} prescribed minimal-match level.

\subsection{Constant Match Contour-Lines\\in Tanaka-Tagoshi Coordinates}

~~The match contour-lines $\gamma_{\Gamma}(\tilde{G})$  
in the (dimensionless) Tanaka-Tagoshi (spin-free) 
waveform parameter-space coordinates $(x_1,x_2)$ have
different geometrical features, depending on the value of $\Gamma$.
In Fig. 1 the inverse of the curvature radius $\rho_\gamma$
of $\gamma_{\Gamma}(\tilde{G})$ is displayed   as a function 
of the polar angle $\phi$ around $\tilde{G}$
for several  representative values of $\Gamma$, for LIGO-I  at 2.5PN order.
The contour-lines $\gamma_{\Gamma}$
are also shown in the insets, where 
$\delta x_i = x_i^{(\tilde{G})}-x_i^{(\tilde{H})}$.

In the limit as  $\Gamma \rightarrow 1$
the contour-lines $\gamma_{\Gamma}$ are asymptotically circular
and $\rho_\gamma^{-1}$ is almost constant.
See, e.g., the $\Gamma=0.99$ contour-line in Fig. 1 top-left.

In a range $\Gamma^* \leq \Gamma < 1$, 
the contour-lines $\gamma_{\Gamma}$ deviate from circular,
but still remain {\em globally convex}. 
For LIGO-I,  $\Gamma^*\approx 0.9225$.
Correspondingly, $\rho_\gamma^{-1}$ remains {\em non negative}
throughout a full rotation around $\tilde{G}$.
See, e.g., the $\Gamma=0.93$ contour-line in Fig. 1 top-right.

For $\Gamma < \Gamma^*$, 
the $\gamma_{\Gamma}$ contour-lines are {\em no longer globally convex}.
In a range  $\Gamma^{\dag} < \Gamma < \Gamma^*$,  
they consist simply of two convex \cite{convex} arcs 
and  two concave ones  joining smoothly, 
to form a "peanut" shape.
See e.g. the $\Gamma=0.914$
contour-line in Fig. 1 bottom-left.
For LIGO-I, $\Gamma^\dag\approx 0.9109$.
As $\Gamma$ is  decreased below $\Gamma^{\dag}$,
the contour-lines become more and more
complicated, as more and more bumps and dents do appear. 
Correspondingly,  the boundary curvature $\rho_\gamma^{-1}$  exhibits more and more
turning points throughout a full rotation around $\tilde{G}$.
See, e.g., the $\Gamma=0.875$  contour-line
in Fig. 1 bottom-right.

For all values of $\Gamma$, the surface $S_{\Gamma}(\tilde{G})$ has always
a centre of symmetry in $\tilde{G}$.

\section{Optimum Template Placement}

~In the following we shall discuss 
the optimum template placement
strategy for each of the above mentioned
minimal-match ranges. Optimum here and henceforth 
means that  requirements {\bf (a)} , {\bf (b)}  and {\bf (c)}
discussed in Sect. 1 will be strictly fulfilled \cite{spill_over}.

\subsection{The Asymptotic $\Gamma \rightarrow 1$ Limit}

~In the asymptotic limit $\Gamma \rightarrow 1$,
the match contour-lines are circular.
Rotational symmetry of the match contour-lines
implies that the templates  should  sit
at  the vertices  of {\em regular} polygons
in the waveform parameter-space.
These regular (open) polygons should besides make up 
a  (regular) {\em tiling}  of  the (Tanaka-Tagoshi) plane \cite{tiling}, 
and  hence can  only be triangles, squares or hexagons.
Let $U$ be one such polygon, and $\{\tilde{G}_i\}$ its vertices.
The  sparsest  templates satisfying (\ref{eq:min_mat1})  
are such that the circles  $\gamma_{\Gamma}(\tilde{G}_i)$  
touch at  a {\em single} point $P$, 
which is obviously the center of symmetry of $U$.
This means that  the curve $\gamma_{\Gamma}(P)$ goes
through all the $\tilde{G}_i$'s.
Hence, the templates should be located 
at the vertices of  a  regular 
triangle, square or hexagon
inscribed in a circle $\gamma_{\Gamma}$.

The obvious question is which polygon is {\em best}  
in terms of  the total number of templates 
needed to cover the admissible-waveform parameter-space 
${\cal T}$ in the Tanaka-Tagoshi plane.
Clearly, the best polygon will be the one for which
the {\em span} of each (and any) template has the largest
measure (area) \cite{tmpt_number}. 

The span (or, in more technical language, the Voronoi set \cite{Voronoi})
of $\tilde{G}_i$ is the set of points $\tilde{H}$
for which $M(\tilde{h},\tilde{g}_i) > M(\tilde{h},\tilde{g}_j),~\forall j \neq i$,
which are the images in the Tanaka-Tagoshi plane of those
waveforms $h$ which will be detected using template $g_i$. 
The Voronoi sets are shown shaded in Fig. 2, 
for the (regular)  triangular, square and hexagonal 
tilings of the (Tanaka-Tagoshi) plane \cite{Voronoi1}.

A choice can be made on the basis of a comparison
among the measures (areas) of the pertinent Voronoi sets. 
Let $\mu(\cdot)$ denote the measure (area), 
$U^{(p)}$ the $p$-gonal tiling-cell,
and $V^{(p)}$  the corresponding Voronoi set.
It is readily seen that  \cite{square_tile}:
\beq
\frac{\mu[V^{(3)}]}{\mu[V^{(4)}]} = 1.299037,~~~~
\frac{\mu[V^{(6)}]}{\mu[V^{(4)}]} = 0.649519,
\label{eq:tile_comp}
\eeq
The triangular tiling is thus seen to be the best choice,
yielding a gain in terms of template-span  of $\sim 30\%$, 
whereas the hexagonal one features a loss of $\sim 35\%$, 
w.r.t. the simples square tiling \cite{misunderstand}.

The triangular {\em tiling} yields a template {\em lattice} 
whose fundamental domain (lattice-cell) is a parallelogram \cite{lattice}. 
Only three vertices of  any  lattice-cell belong to a single
curve  $\gamma_{\Gamma}$.

\subsection{The  Range  $\Gamma > \Gamma^*$}

In the  range $\Gamma > \Gamma^*$,
the contour-lines $\gamma_{\Gamma}(\tilde{G})$
are no longer circles,  but still {\em globally-convex}
and center-symmetric.
Circular symmetry being lost, 
we shall still seek the sparsest template collocation
subject to (\ref{eq:min_mat1}) 
among  the non-regular, but still center-symmetric
(open) $p$-gons ($p=3,4,6$)  with all vertices on  $\gamma_{\Gamma}$,
which tile the Tanaka-Tagoshi plane.

It makes sense to compare for this case
the template-span (Voronoi set) measures  (areas)
$\mu[V^{(p)}]$, $p=3,4,6$
to the template-span measure (area)  $\mu[V_0^{(4)}]$ 
of the square tiling  \cite{square_tile}
pertinent to the naive (but usual) approximation of circular match contour-lines.
The ratios
\beq
r_p = \frac{\mu[V^{(p)}]}{\mu[V_0^{(4)}]},~~~p=3,4,6.
\label{eq:tile_gains}
\eeq
are displayed in Fig. 3 as functions of $\Gamma$  for LIGO-I.
Not unexpectedly, the triangular tiling is once more the best.
This conclusion is valid for all first generation antennas, and
the pertinent values of $r_3$ have been collected in Table-I.

\begin{center}
\begin{tabular}{|l|c||}
\hline\hline
Antenna &  $r_3$\\
\hline
TAMA300  &   1.58 \\
\hline
LIGO-I & 1.43\\
\hline
GEO600   &    1.58\\
\hline
VIRGO & 1.38\\
\hline\hline
\end{tabular}
$$~$$
{\em Table I.}- The ratio $r_3$, eq. (\ref{eq:tile_gains}), for first generation 
interferometers at $\Gamma=0.97$.
\end{center}
 
\subsection{The  Range $\Gamma < \Gamma^*$}

By an argument of  continuity 
(i.e., since the match contour-lines 
corresponding to $\Gamma < \Gamma^*$ 
can be obtained from those 
corresponding to  $\Gamma > \Gamma^*$
by continuous transformations) 
we shall still seek the optimum tiling of Tanaka-Tagoshi plane
using the (open) triangles with (all) vertices on $\gamma_{\Gamma}$.
However, the match contour-lines being no longer globally convex,
the {\em largest-area} triangle inscribed in $\gamma_{\Gamma}$
does {\em not} necessarily cope with eq.  (\ref{eq:min_mat1}), 
as e.g. exemplified in Fig. 4.

One should accordingly seek three points $\tilde{G}_i$ 
on $\gamma_{\Gamma}$  yielding the largest-area 
triangle $\tilde{G}_0\tilde{G}_1\tilde{G}_2$,
subject to eq. (\ref{eq:min_mat1}).
The minimal-match condition (\ref{eq:min_mat1}) is equivalent to
\beq
C-\bigcup_i\left[ C \cap S_{\Gamma}(\tilde{G}_i) \right] =\emptyset,
\label{eq:min_mat2}
\eeq
where $C$ is the (parallelogram)  fundamental-domain of the lattice
corresponding to a given triangular tiling.
Condition (\ref{eq:min_mat2}) is most easily checked,
involving only  four patches $S_{\Gamma}$.

Let  $\tilde{P},\tilde{P'},\tilde{Q},\tilde{Q'}$ 
the points (ordered,  e.g., counterclockwise)
where $\gamma_{\Gamma}$ and its convex-hull  \cite{CH}
detach, shown in Fig. 5.
One can readily prove that  two vertices of the sought optimum triangular-tile,
say $\tilde{G}_0,~\tilde{G}_1$,  
should be sought  on the arc $\tilde{P}\tilde{P'}$ 
(or $\tilde{Q}\tilde{Q'}$)  of  $\gamma_{\Gamma}$,
while the third should be sought on the arc $\tilde{Q}\tilde{Q'}$ 
(resp. $\tilde{P}\tilde{P'}$).
It is also readily proved that the lattice vector $\tilde{G}_0\tilde{G}_1$,
after a trivial translation making its midpoint coincident with $\tilde{G}$,
should be contained in the butterfly-shaped region  
$\left(\tilde{P}\tilde{G}\tilde{Q'} \cup  \tilde{Q}\tilde{G}\tilde{P'}\right) \cap S_{\Gamma}(\tilde{G})$ depicted in Fig. 5.  

The maximum-area triangular tiles subject to the  constraint (\ref{eq:min_mat1}),
are shown in Fig. 6,  for several values of $\Gamma$
in the range $0.7 \leq \Gamma < 1$ (note that different scales
are used in the the figure, for better readability).
The lattice-cell measure corresponding to the optimum triangular-tile 
is displayed in Fig. 7 as a function of $\Gamma$,
together with the number of templates needed to cover the 
companion-mass range 
$0.2M_{\odot} \leq m_1\leq m_2 \leq 10 M_{\odot}$,
for a number of representative minimal-match values.

\section{Correlator Bank Economization via Cardinal Interpolation }

In  recent papers it has been shown that a sensible reduction
($\approx 75\%$ for 1PN and higher-PN-order templates)
in the number of correlators to be computed in order
to achieve  a prescribed  minimal-match
can be obtained using cardinal interpolation,
thanks to the quasi-band-limited property 
of the match function \cite{Card0}, \cite{Card1}.
These results were obtained under the simplest
assumption of a square-cell template-lattice.

Two obvious questions are now in order: 
i) whether/to what extent cardinal interpolation is still effective
when using the optimum triangular-tiling discussed in the previous
sections \cite{skew_grid}, and 
ii)  whether/to what extent  cardinal interpolation 
is  still  useful at relatively low $\Gamma$ values, 
as needed in the early step(s) of hierarchical searches.
Numerical simulations have been run to clarify both issues.

In Fig.  8 the template density reduction  \cite{gain_def} 
obtained by cardinal interpolation among optimally-placed templates
is displayed as a function of the prescribed minimal-match $\Gamma$. 
In the inset of Fig. 8, the (boosted) minimal-match values obtained
after cardinal interpolation among the optimally-placed templates
corresponding to a number of representative values of $\Gamma$ 
are also listed.

At $\Gamma=0.97$ a template density reduction
$\approx 2.79$ is obtained.
The apparent discrepancy between this value and the one
reported in \cite{Card1} is readily explained.
The template density required 
to achieve a prescribed minimal-match
when using cardinal interpolation
is essentially independent from 
the {\em shape} of the lattice fundamental-domain,
depending only on its  {\em measure} (area), 
according to the well-known 
Nyquist-rate condition \cite{2D_Nyquist_rate}.

Thus, the template density reduction  
$\approx 4$  in \cite{Card1}, \cite{TaTa1} 
when using the simplest square-cell lattice 
is seen to result from  two factors:
a factor $\approx 2.79$, expressing the template density reduction
due to cardinal interpolation when using the {\em optimum} triangular-tiling,
times a factor $\approx 1.43$ expressing the template density reduction
implied in using the optimum  triangular-tiling
instead of the simplest square one.

\section{Conclusions}

We presented a simple and systematic  procedure 
for constructing the sparsest  lattice of templates 
subject to a given minimal-match constraint
in the range between  $\Gamma \sim 1$ 
down to $\Gamma \approx 0.7$ (and below) 
in the (spin-free) Tanaka-Tagoshi parameter-space
of post-1PN gravitational wave chirps.

We also showed that cardinal interpolation can be effective
for correlator-bank economization both
in the late and early stage(s) of multi-step
hierarchical searches.

\section*{Acknowledgements}

This work has been sponsored in part by the EC
through a senior visiting scientist grant to I.M. Pinto at NAO, Tokyo, JP.
I.M. Pinto wishes to thank the TAMA staff at NAO, 
and in particular prof. Fujimoto  Masa-Katsu and prof. Kawamura Seiji 
for gracious hospitality and stimulating discussions.


\newpage


\begin{center}
{\bf CAPTIONS TO THE FIGURES}
\end{center}
$$~$$
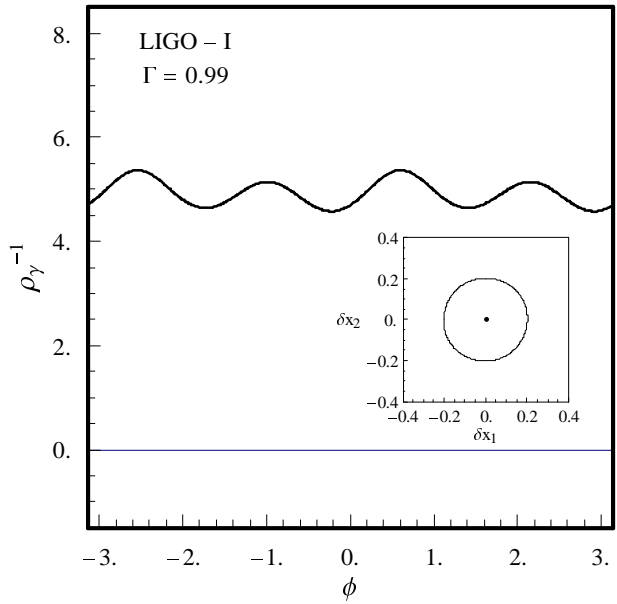
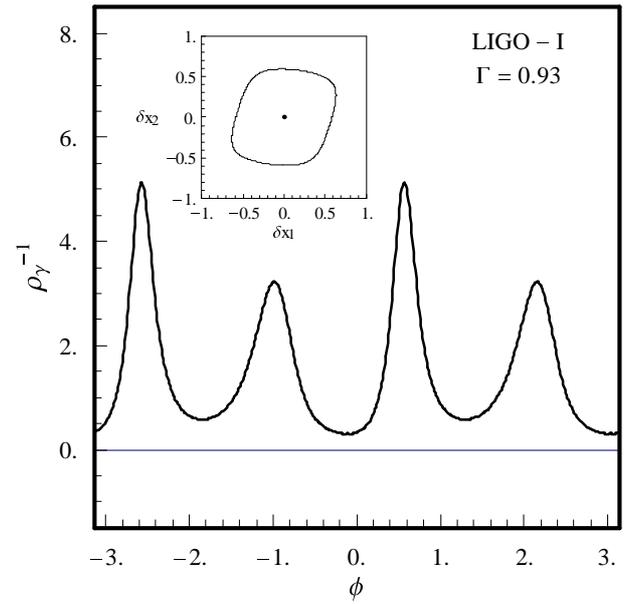
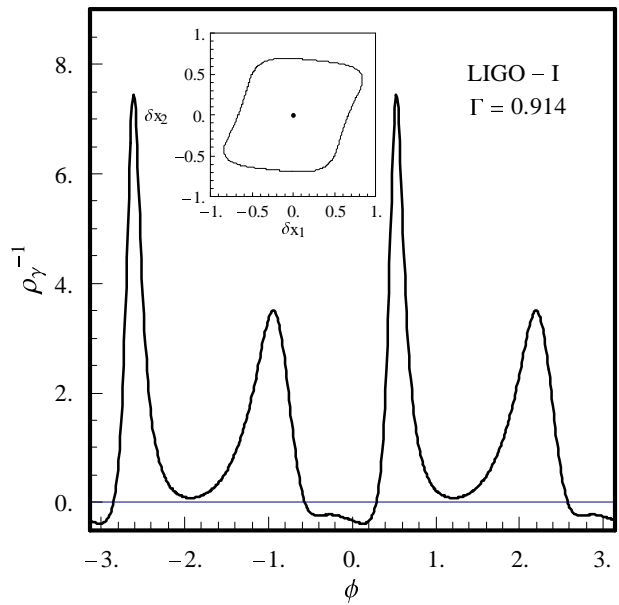
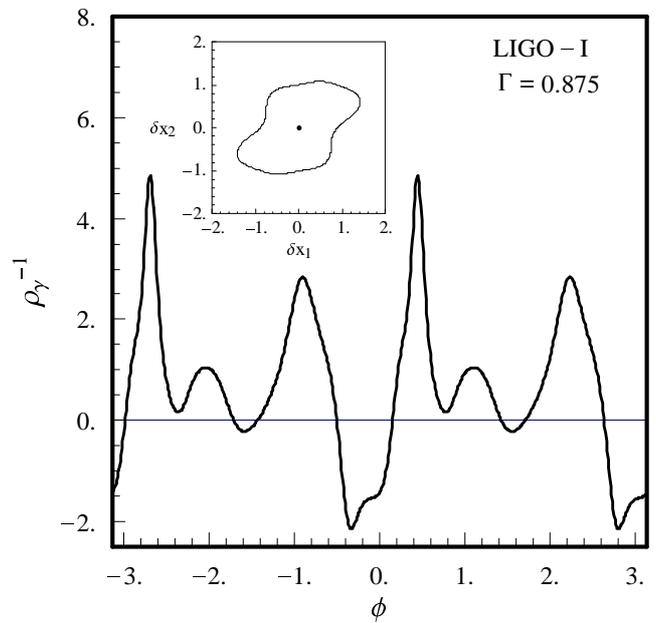
Fig 1 -  Match contour-lines in Tanaka-Tagoshi coordinates. 
Curvature $\rho_{\gamma}^{-1}$  of   $\gamma_{\Gamma}$
vs.  polar angle $\phi$  (LIGO-I, 2.5PN).\\
$$~$$
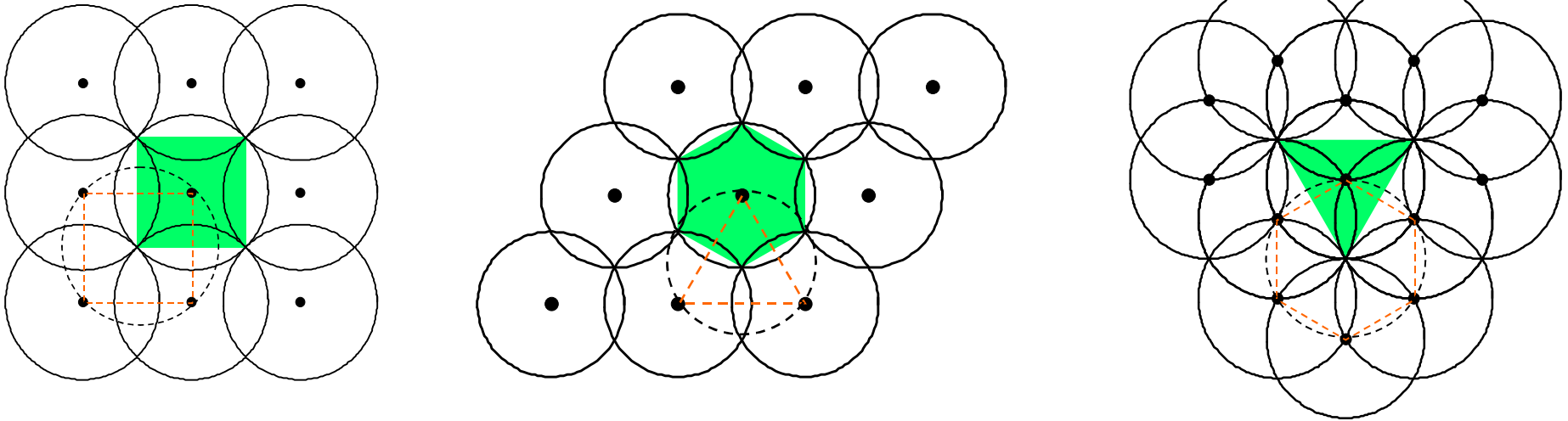
Fig. 2 - Triangular, square and hexagonal tilings and  pertinent Voronoi sets.\\
$$~$$
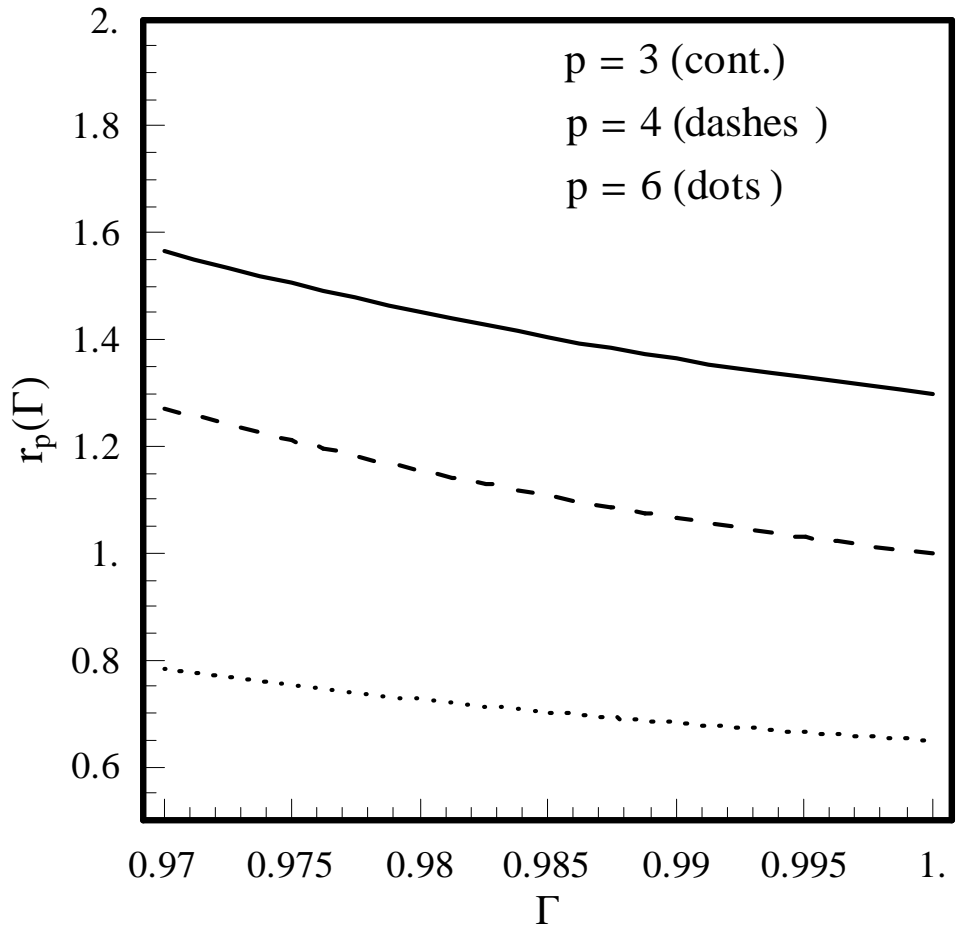
Fig. 3 - The ratios $r_p$, $p=3,4,6 $ relevant to Eq. (\ref{eq:tile_gains}) as functions of $\Gamma$ (LIGO-I, 2.5PN).\\
$$~$$
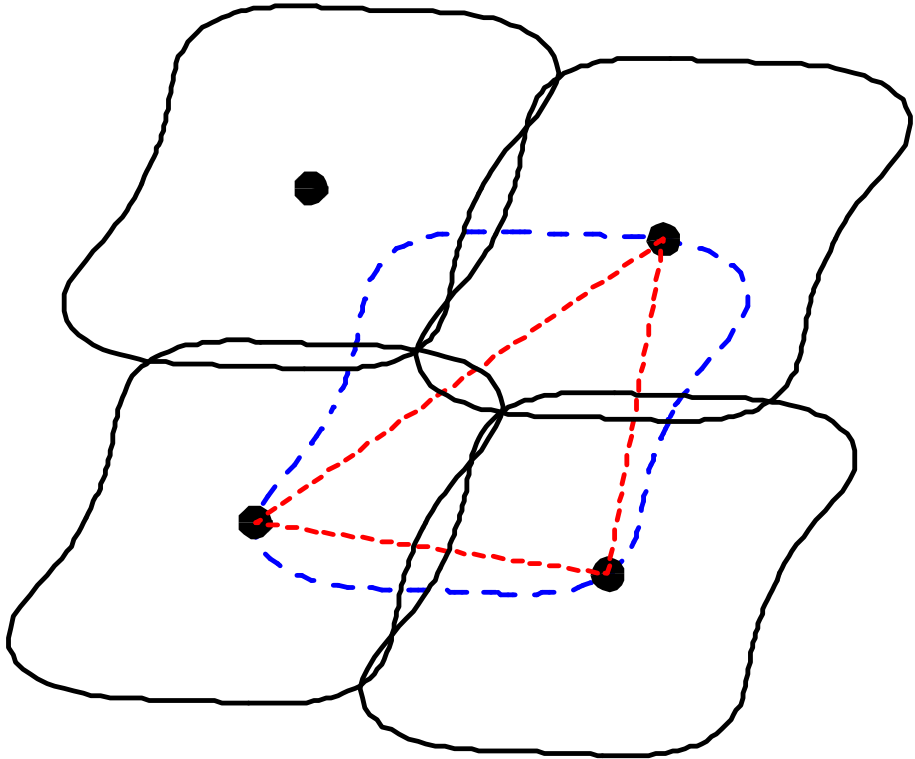
Fig. 4 - Inscribed triangle with largest area does not necessarily cope with minimal-match condition for $\Gamma < \Gamma^*$  (LIGO-I, 2.5PN, $\Gamma=0.9$).\\
$$~$$
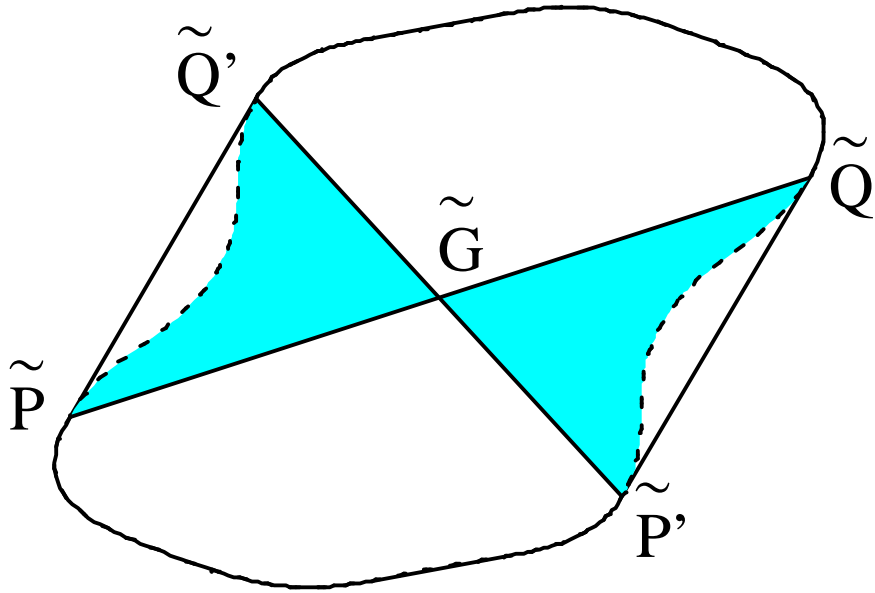
Fig. 5 - The butterfly-shaped set $\left(\tilde{P}\tilde{G}\tilde{Q'} \cup  \tilde{Q}\tilde{G}\tilde{P'}\right) \cap S_{\Gamma}(\tilde{G})$ (LIGO-I, 2.5PN,  $\Gamma=0.875$).\\
$$~$$
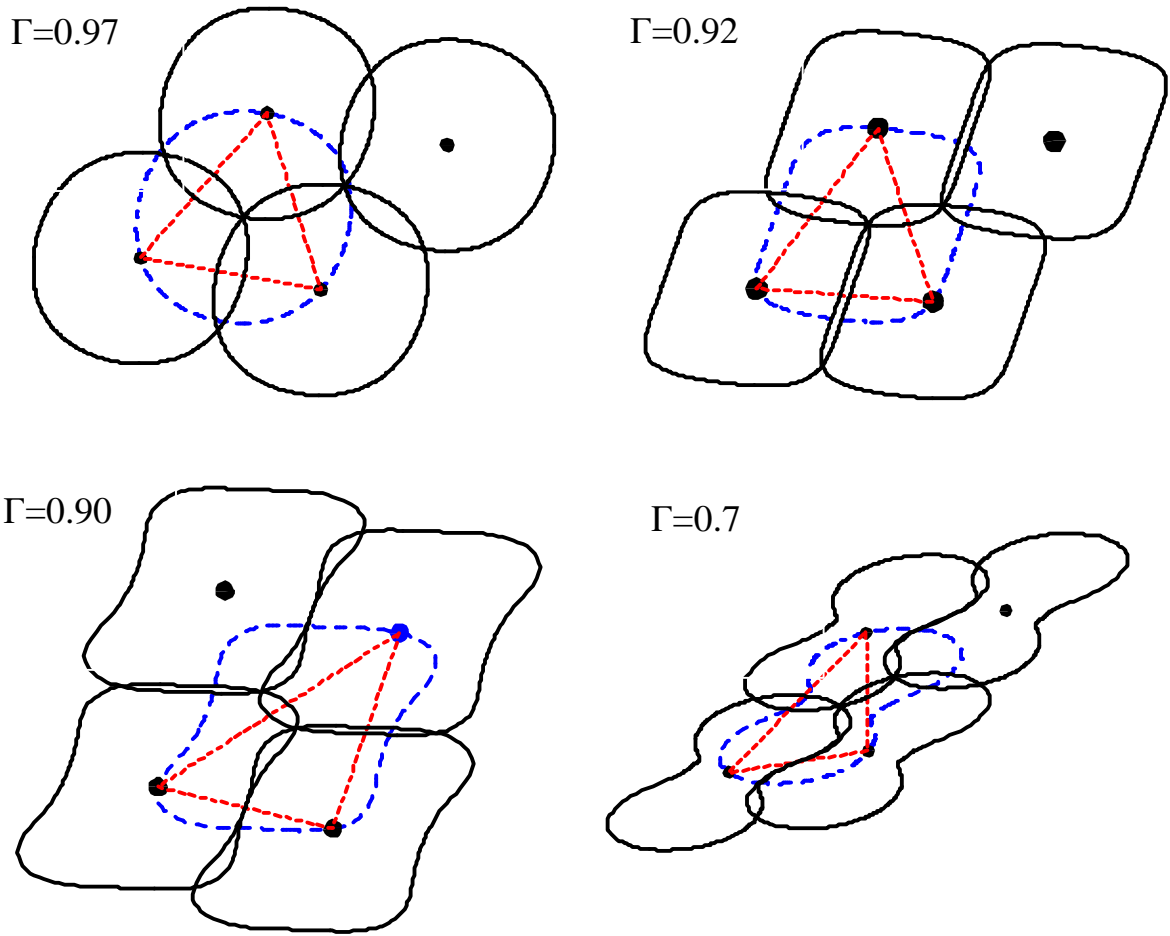
Fig. 6 - Optimum triangular-tilings and template-lattices  for several values of $\Gamma$ (LIGO-I, 2.5PN).\\
$$~$$
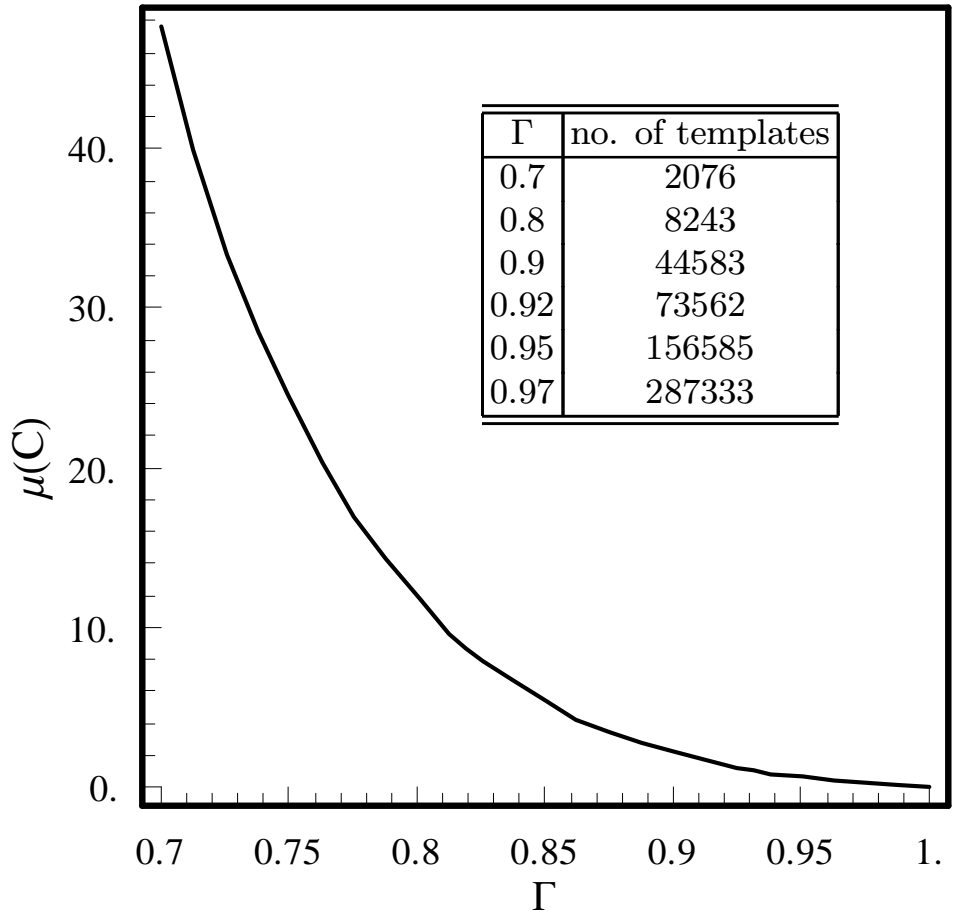
Fig. 7 - Lattice-cell area corresponding to optimum triangular-tiling and number of templates needed to cover the range $0.2 M_{\odot} \leq m_1 \leq m_2 \leq 10 M_{\odot}$ vs. minimal-match (LIGO-I, 2.5PN).\\
$$~$$
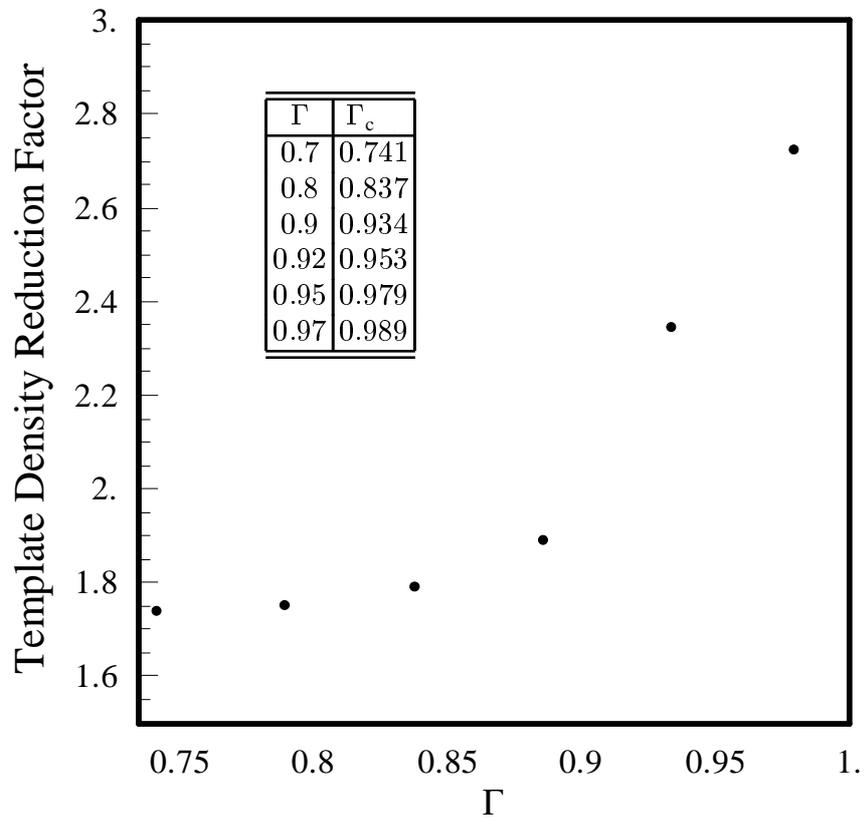
Fig. 8 - Cardinal-interpolation gain (template density reduction factor and minimal-match boost) for optimum triangular-tiling (LIGO-I, 2.5PN).\\



\begin{thebibliography}{99}
\bibitem{sources}{B.F. Schutz, Class. Quant. Grav. {\bf 16},  A131 (1999).}
\bibitem{signals}{T. Damour, B.R. Iyer and B.S. Sathyaprakash,
Phys. Rev. {\bf  D63}, 044023 (2001).}
\bibitem{Helstrom}{C.W. Helstr\"{o}m, {\em Statistical Theory of Signal Detection} (Pergamon Press, Oxford, 1968).}
\bibitem{Sat_Dhu}{B.S. Sathyaprakash and S.V. Dhurandhar, Phys. Rev. {\bf D44}, 3819 (1991).}
\bibitem{Dhu_Sat}{S.V. Dhurandhar and B.S. Sathyaprakash, Phys. Rev. {\bf D49}, 1707 (1994).}
\bibitem{Apo}{T.A.  Apostolatos, Phys. Rev. {\bf D52}, 605 (1995).}
\bibitem{Owen}{B.J. Owen, Phys. Rev. {\bf D53}, 6749 (1996).}
\bibitem{Owe_Sat}{B.J. Owen and B.S. Sathyaprakash, Phys. Rev. {\bf D60}, 022002 (1999).}
\bibitem{partialmax}{The maximization is trivially accomplished
by taking the largest absolute element in the (inverse, fast) DFT 
of $h(f)g^*(f)/\Pi(f)$ \cite{Owen}.}
\bibitem{Taylor}{J.H. Taylor, Rev. Mod. Phys., {\bf 66}, 711 (1994).}
\bibitem{GRASP}{GRASP is the current {\em de facto} standard
software package for gravitational wave data analysis, available from
http://www.lsc-group.phys.uwm.edu/$\sim$ballen/grasp-distribution.}
\bibitem{falsealarm}{It should be noted that template {\em overpopulation}
even in a {\em subset} of the waveform parameter-space increases 
the false-alarm probability of the {\em whole} bank, at a fixed
detection threshold level.}
\bibitem{MoDhu}{S.D. Mohanty and S.V. Dhurandhar, Phys. Rev.  {\bf D54}, 7108 (1996).}
\bibitem{Mohanty}{S.D. Mohanty, Phys. Rev. {\bf D57}, 630 (1998).}
\bibitem{alternative}{Multi-step searches could be implemented following different routes.
One might alternatively use lower PN-order templates
in the early stage(s)  to obtain a coarse estimate
of (a subset of)  candidate-event source parameters, and
then use the best-available higher PN-order templates for
a refined high-$\Gamma$ search  in the neighbourhood(s)
of the candidate source(s) sieved in the first step.
No quantitative analysis of this procedure has been given yet.
See  A.E. Chronopoulos and T.A.  Apostolatos, Phys. Rev. {\bf  D64}, 042003 (2001)
for an updated discussion of the problem of  matching waveforms of different PN orders.}
\bibitem{TaTa}{T. Tanaka and H. Tagoshi, Phys. Rev. {\bf D62}, 082001 (2000).}
\bibitem{TaTa1}{R.P. Croce, Th. Demma, V. Pierro and I.M. Pinto, Phys. Rev. {\bf D64}, 042005  (2001).}
\bibitem{lattice}{A $2D$ (regular) lattice is the set  of intersections
between  two (non parallel) sets of parallel equispaced lines. 
The resulting  checkerboard  tiles the plane with (equal) parallelograms.
Any of these parallelograms is the lattice {\em fundamental-cell}. 
The two vectors from any grid-node to its nearest-neighbor fundamental-cell vertices
are the {\em lattice vectors}.}
\bibitem{spill_over}{The template spill-over condition {\bf d)}
will be approximately satisfied at the same level of accuracy 
provided by current placement algorithms.}
\bibitem{TaTa2}{Basically, the Tanaka-Tagoshi transformation 
is obtained by requiring  ${\cal T}$ to touch $\Sigma$ 
at its three vertices. See Appendix-B of  \cite{TaTa1} for details.}
\bibitem{convex}{Convex (concave) arc here means that the (open) segment  
between the arc endpoints is completely included in (respectively, external to)  $S_{\Gamma}$.}
\bibitem{tiling}{A (polygonal) tiling of  $R^2$ is a countable family 
of (open) polygons $U_i$, which i) are disjoint and ii) are such 
that their closures cover $R^2$.
In a {\em regular}  tiling all tiles $U_i$ are congruent, i.e., 
can be made to coincide up to suitable translations and rotations.}
\bibitem{tmpt_number}{The total number of templates needed
to cover ${\cal T}$ is approximately equal to the ratio 
between the measure (area) of $\Sigma$ 
and the measure (area) of the template {\em span}.} 
\bibitem{Voronoi}{Given a set of points ${\cal G} \subseteq R^n$, 
the Voronoi set or {\em proximity locus}  of $G_k \in  {\cal G}$
is the set  $V(G_k) = \{ x \in R^n : \forall  h \neq k, D(x,G_h) > D(x,G_k)\}$,
where $D(\cdot,\cdot)$ is a suitable distance (metric).}
\bibitem{Voronoi1}{It is seen from Fig. 2 that, but for trivial
translations, $U^{(3)}\!=\!V^{(6)}$,  $U^{(6)}\!=\!V^{(3)}$, and
$U^{(4)}\!=\!V^{(4)}$, i.e., the triangular and hexagonal tiling
are Voronoi-dual, while the square tiling is Voronoi self-dual.}
\bibitem{square_tile}{We remind that for circular match contour-lines, 
the inscribed square-tile area is given by 
$\mu[V^{(4)}]=2(1-\Gamma)$ .}
\bibitem{misunderstand}{Occasional confusion is made in the technical
Literature between the tiling-cell and tiling-cell's span (Voronoi set) measures, 
leading  e.g.  to the often quoted  statement  that  hexagonal tiling would be
the most efficient.}
\bibitem{CH}{The  convex hull of a set  $U$  is the smallest
 convex  set containing  $U$. }
\bibitem{Card0}{R.P. Croce, Th. Demma, V. Pierro, I.M. Pinto and F. Postiglione, Phys. Rev. {\bf D62}, 124020 (2000).}
\bibitem{Card1}{R.P. Croce, Th. Demma, V. Pierro, I.M. Pinto, D. Churches and B.S. Sathyaprakash, Phys. Rev. {\bf D62}, 121101(R) (2000).}
\bibitem{skew_grid}{The standard  $2D$ cardinal-interpolation formula
applies to the case where the interpolation points form a rectangular grid.
The cardinal-interpolation formula
for the case where the interpolated points form a skew grid 
is more or less obviously obtained 
by first applying a trivial coordinate transformation 
which brings the grid into a rectangular one,
then writing down the usual (rectangular-grid) cardinal-interpolation formula, 
and finally switching back to the original coordinates.}
\bibitem{gain_def}{The template-density reduction factor is defined here as
the ratio between the areas of the (optimal) triangular-tiles yielding
the same minimal-match with and without cardinal interpolation.}
\bibitem{2D_Nyquist_rate}{See, e.g., J.R. Higgins, {\em Sampling Theory in Fourier and Signal Analysis} (Clarendon Press, Oxford, 1996), ch. 14.}

\end{thebibliography}
\end{document}